\documentclass[12pt]{iopart}

\usepackage{iopams}
\usepackage{braket}
\usepackage{amsfonts}
\usepackage{color}
\usepackage[dvipsnames]{xcolor}
\usepackage{amssymb}
\usepackage{mathrsfs}
\usepackage{graphicx}
\usepackage{lmodern}
\usepackage{lineno}
\usepackage{hyperref}
\usepackage[normalem]{ulem}
\usepackage{soul}

\newcommand{\beq}{\begin{eqnarray}}
\newcommand{\eeq}{\end{eqnarray}}

\newcommand{\sinc}{\mathrm{sinc}}
\newcommand{\Imp}{\mathrm{Im}\,}
\newcommand{\Ai}{\mathrm{Ai}\,}
\newcommand{\Exp}{\mathrm{Exp}_{\star}}
\newcommand{\nn}{\nonumber}
\newcommand{\blue}[1]{\textcolor{blue}{#1}}

\def\keywords#1{\vspace{10pt}
     \begin{indented}
     \item[]\rm Keywords: #1\par
     \end{indented}}

\begin{document}



\title{Star exponentials from propagators and path integrals}
\author{Jasel Berra--Montiel$^{1,2}$, Hugo Garc\'ia-Compe\'an$^{3}$ and Alberto Molgado$^{1}$}

\address{$^{1}$ Facultad de Ciencias, Universidad Aut\'onoma de San Luis 
Potos\'{\i} \\
Campus Pedregal, Av. Parque Chapultepec 1610, Col. Privadas del Pedregal, San
Luis Potos\'{\i}, SLP, 78217, Mexico}
\address{$^2$ Dipartimento di Fisica ``Ettore Pancini", Universit\'a degli studi di Napoli ``Federico II", Complesso Univ. Monte S. Angelo, I-80126 Napoli, Italy}
\address{$^3$Departamento de F\'isica, Centro de Investigaci\'on y de Estuios Avanzados del IPN \\
P.0. Box 14-740, Ciudad de M\'exico, 07000, Mexico}

\eads{\mailto{\textcolor{blue}{jasel.berra@uaslp.mx}},\ 
\mailto{\textcolor{blue}{hugo.compean@cinvestav.mx}}\
\mailto{\textcolor{blue}{alberto.molgado@uaslp.mx}} 
}


\begin{abstract}
In this paper we address the relation between the star exponentials emerging within the Deformation Quantization formalism and Feynman's path integrals associated with propagators in quantum dynamics.    In order to obtain such a relation, we start by visualizing the quantum propagator as an integral transform of the star exponential by means of the symbol corresponding to the time evolution operator and, thus, we introduce Feynman's path integral 
representation of the propagator as a sum over all the classical histories.  The star exponential thus constructed  has the advantage that it does not depend on the convergence of formal series, as commonly understood within the context of Deformation Quantization.  We include some basic examples to illustrate our findings, recovering standard results reported in the literature. Further, for an arbitrary finite dimensional system, we use the star exponential introduced here in order to find a particular representation of the star product which resembles the one encountered in the context of the quantum field theory for a Poisson sigma model.
\end{abstract}

\keywords{Deformation quantization, star product, path integrals}
\ams{81S30, 46F10, 53D55, 81S40}


\section{Introduction}

Deformation Quantization stands for an implementation of quantum mechanics in the classical phase space realized by introducing a non-commutative associative product that replaces the standard pointwise product~\cite{Curtright,Teaching,Cosmas}.  Such product, commonly denoted as the Moyal star product, is obtained by means of an appropriate convolution of the Weyl's quantization map that associates classical observables with symmetric quantum operators acting on elements of a Hilbert space.   The Moyal star product is particularly useful to investigate time evolution  for quantum states by generalizing Liouville's theorem of classical mechanics for an appropriate probability distribution which corresponds to the density operator in quantum dynamics, namely the Wigner distribution.   The formal solution for the time evolved Wigner distribution is thus obtained through the star exponential of the classical Hamiltonian, in complete analogy to the construction of the unitary operators associated with the time evolution of the density operator~\cite{Curtright,Bayen}.  Nevertheless, the relevance of the star exponential functions has been underestimated throughout the years basically due to two main issues:  on the one hand,  the issue of convergence of star exponential functions remains so far an open problem while, on the other hand, its construction for physically motivated examples has been explored barely in the literature.    The situation may be even worse for gauge systems for which, following a recent proposal introduced in reference~\cite{SGA}, in order to obtain a gauge-invariant Wigner distribution one may implement the group averaging technique over the symmetry group by means of star exponential functions of the constraint generators.    This last proposal relies on previous developments that appropriately construct gauge invariant states in a physical Hilbert space.    

Bearing this in mind, the main aim of the present paper is to review a relation between the star exponentials defined in phase space and the well-known path integrals associated to propagators in quantum dynamics.    Although the path integral representation of quantum dynamics has been realized in the star product formalism of Deformation Quantization, see for example~\cite{Curtright},~\cite{Sharan},~\cite{Samson},   we introduce here the star exponential of the classical Hamiltonian function by means of a Fourier transform of the quantum mechanical propagator which provides us with an authentic manner to obtain star exponentials without depending on the issue of convergence of the formal series involved in the Moyal star product formalism.  In particular, our developments may become valuable by considering Feynman's path integral representation of the propagator as a sum over all the classical histories in phase space.  The subject under study here has been scarcely developed in references~\cite{Teaching,Hirshfeld02,Yoshioka,Wong,Takabayasi,Dittrich}, where the relation 
between the star exponential function and path integrals was considered 
either for reduced symmetry models or under limited physical situations.   Our intention is thus to explore the star exponential through path integrals  for specific physically motivated examples where we may easily appreciate the relevance of our construction.   Undoubtedly, the primary motivation for this work is directed towards its natural extension to field theories for which Feynman's path integral formulation has been explored extensively throughout the years,  including a well-defined version for the case of gauge theories.

The paper is organized as follows. In Section~\ref{sec:DQ} we briefly review the concept of the 
star product within the Deformation Quantization program with the aim of introducing the star exponential functions in phase space.  In Section~\ref{sec:StarExp} we explore the relation between such star exponential functions and  
the propagators in quantum mechanics and, also, the induced 
natural association of the former through Feynman's path integral representation of the latter.   We also include in this section some simple examples to illustrate our construction of the 
star exponential functions and, within our setup, we also introduce a star product 
representation in terms of path integrals which matches the one previously 
discussed for a Poisson sigma model.   Finally, we include some concluding remarks and perspectives in Section~\ref{sec:conclu}.

\section{Deformation quantization and star products}
\label{sec:DQ}

Quantization refers to the process of passing from a classical mechanical system, generally described by a Poisson structure, into its corresponding quantum system. A precise mathematical description for such a procedure must be not only consistent with the axioms of quantum mechanics, but it also needs to provide a correspondence $\mathcal{Q}:f\mapsto \mathcal{Q}(f)$, mapping real valued functions $f$ on the classical phase space, denoted by $\Gamma$, into self-adjoint operators $\mathcal{Q}(f)$, defined on the quantum Hilbert space $\mathcal{H}$. The main difference between classical and quantum frameworks stands on the algebraic structure of the observable algebra. On the one hand, a classical system is characterized by a set of commutative smooth functions on a Poisson manifold, whilst a quantum system is described by a non-commutative algebra of operators. In order to construct a fiducial correspondence map, $\mathcal{Q}$, we must impose the following properties:
\begin{enumerate}
\item $\mathcal{Q}(1)=\widehat{I}$, $\mathcal{Q}(x)=\widehat{X}$ and $\mathcal{Q}(p)=\widehat{P}=-i\hbar\frac{\partial}{\partial x}$.
\item $f\mapsto \mathcal{Q}(f)$ is linear.
\item $[\mathcal{Q}(f),\mathcal{Q}(g)]=i\hbar\mathcal{Q}(\{f,g\})$.
\item For any $\phi:\mathbb{R}\to\mathbb{R}$, $\mathcal{Q}(\phi\circ f)=\phi(\mathcal{Q}(f))$,
\end{enumerate}           
where $\widehat{X}$ and $\widehat{P}$ are the usual position and momentum operators, respectively. 
Property (iv) is known as the von Neumann rule, which guarantees that polynomial functions are mapped into polynomial operators \cite{vonNeumann}. Nevertheless, according to Groenewold's "no-go" theorem, there does not exist a linear map that takes the Poisson algebra $C^{\infty}(\Gamma)$ into the Lie algebra of the corresponding operators \cite{Groenewold}. In order to circumvent this shortcoming, it is conceivable to assume that condition (iii) is fulfilled only asymptotically in the limit $\hbar\to 0$, this quantization approach is commonly denominated as deformation quantization. The theory of deformation quantization, first introduced in the seminal papers \cite{Bayen}, consists in a deformation of the algebraic structure of classical observables by replacing the usual pointwise product of functions with a noncommutative and associative one. This noncommutative product, usually termed as the star product, allows to define the quantum algebra of observables as a deformation of the Poisson structure, inherent to the classical phase space, in terms of formal power series, but, without the need to radically change the nature of observables by employing operators on a Hilbert space.  

The origins of the deformation quantization approach can be traced back to the Weyl's quantization map, which associates to any classical observable, $f(\mathbf{x},\mathbf{p})\in L^{2}(\mathbb{R}^{2n})$ defined on the phase space $\Gamma=\mathbb{R}^{2n}$, an operator (the quantum observable) acting on the Hilbert space $\mathcal{H}=L^{2}(\mathbb{R}^{n})$ by 
\begin{equation}
\mathcal{Q}_W(f)=\frac{1}{(2\pi)^{n}}\int_{\mathbb{R}^{2n}}\widetilde{f}(\mathbf{a},\mathbf{b})e^{i(\mathbf{a}\cdot\mathbf{\widehat{X}}+\mathbf{b}\cdot\mathbf{\widehat{P}})}d\mathbf{a}\,d\mathbf{b}   \,,
\end{equation}
where $\widetilde{f}(\mathbf{a},\mathbf{b})$ denotes the Fourier transform on $\mathbb{R}^{2n}$, and the components of the operators $\widehat{\mathbf{X}}$, $\widehat{\mathbf{P}}$ satisfy the canonical commutation relations $[\widehat{X}_{i},\widehat{P}_{j}]=i\hbar\delta_{ij}$, with $i,j=1,\ldots, n$. Properly speaking, the integral is taken in the weak operator topology and is not necessarily absolutely convergent \cite{Takhtajan}. By explicitly writing the Fourier transform $\widetilde{f}(\mathbf{a},\mathbf{b})$ as
\begin{equation}
\widetilde{f}(\mathbf{a},\mathbf{b})=\frac{1}{(2\pi)^{n}}\int_{\mathbb{R}^{2n}}f(\mathbf{u},\mathbf{v})e^{-i(\mathbf{u}\cdot\mathbf{a}+\mathbf{v}\cdot\mathbf{b})}d\mathbf{u}\,d\mathbf{v}   \,,
\end{equation}
the Weyl's map reads
\begin{equation}\label{Weylmap}
\mathcal{Q}_{W}(f)=\frac{1}{(2\pi\hbar)^{2n}}\int_{\mathbb{R}^{2n}}f(\mathbf{x},\mathbf{p})\widehat{\Omega}(\mathbf{x},\mathbf{p})d\mathbf{x}\,d\mathbf{p}   \,, 
\end{equation}
where the operator $\widehat{\Omega}(\mathbf{x},\mathbf{p})$ corresponds to the Weyl-Stratonovich operator \cite{Stratonovich}, and it is given by
\begin{equation}
\widehat{\Omega}(\mathbf{x},\mathbf{p})=\left(\frac{\hbar}{2\pi}\right)^{n}\int_{\mathbb{R}^{2n}}e^{i(\mathbf{a}\cdot(\widehat{\mathbf{X}}-\mathbf{x})+\mathbf{b}\cdot(\widehat{\mathbf{P}}-\mathbf{p}))}d\mathbf{a}\,d\mathbf{b}  \,. 
\end{equation}
This integral operator may be identified as a Hilbert-Schmidt operator on $L^{2}(\mathbb{R}^{n})$, that is, $\widehat{\Omega}(\mathbf{x},\mathbf{p})\in HS(L^{2}(\mathbb{R}^{n}))$,  since it possesses a well defined trace~\cite{Reed}. In particular the Weyl-Stratonovich operator $\widehat{\Omega}(\textbf{x},\textbf{p})$ turns out to be self-adjoint and satisfies, $\tr(\widehat{\Omega}(\textbf{x},\textbf{p}))=1$, $\tr(\widehat{\Omega}(\textbf{x},\textbf{p})\widehat{\Omega}(\textbf{x}',\textbf{p}'))=2\pi\hbar\delta(\mathbf{x}-\mathbf{x}')\delta(\mathbf{p}-\mathbf{p}')$, regardless of the orthonormal basis. By applying the Weyl quantization map on any $\psi\in L^{2}(\mathbb{R}^{n})$, we obtain
\begin{eqnarray}
\mathcal{Q}_{W}(f)\psi(\mathbf{x})&=&\frac{1}{(2\pi\hbar)^{2n}}\int_{\mathbb{R}^{2n}}f((\mathbf{x}+\mathbf{y})/2,\mathbf{p})e^{-i(\mathbf{y}-\mathbf{x})\cdot\mathbf{p}/\hbar}\psi(\mathbf{y})d\mathbf{p}\,d\mathbf{y}  \,, \\
&=&\int_{\mathbb{R}^{n}}\kappa_{f}(\mathbf{x},\mathbf{y})\psi(\mathbf{y})d\mathbf{y}  \,, \nonumber
\end{eqnarray}
where, $\kappa_{f}(\mathbf{x},\mathbf{y})\in L^2(\mathbb{R}^{2n})$, corresponds to an integral kernel characterizing the Hilbert-Schmidt operator $\mathcal{Q}_{W}(f):=\widehat{F}\in HS(L^{2}(\mathbb{R}^{n}))$ \cite{Reed}. Furthermore, for all $f\in L^{2}(\mathbb{R}^{2n})$ the Weyl's map fulfills $\mathcal{Q}_{W}(\bar{f})=\mathcal{Q}_{W}(f)^{*}$, where $\bar{f}$ denotes the complex conjugate of $f$. Since the integral in the momentum variables, denoted by $\mathbf{p}$, defines a partial Fourier transform, we should, strictly speaking, replace this integral by a limit in the norm topology in order to compute the Fourier transform by applying Fubini's theorem and the Plancherel formula for $\mathbb{R}^{n}$ \cite{Reiter}. Now, by multiplying equation (\ref{Weylmap}) by the operator $\widehat{\Omega}(\mathbf{x},\mathbf{p})$, and taking the trace on both sides, it follows that the inverse of the Weyl's quantization map is given by
\begin{eqnarray}\label{InverseWeyl}
\mathcal{Q}_{W}^{-1}(\widehat{F})(\mathbf{x},\mathbf{p})&=&\tr\bigg(\widehat{\Omega}(\mathbf{x},\mathbf{p})\widehat{F}\bigg)  \\
&=& \hbar^{n}\int_{\mathbb{R}^{n}}\kappa_{f}(\mathbf{x}-\hbar\mathbf{y}/2,\mathbf{x}+\hbar\mathbf{y}/2)e^{i\mathbf{y}\cdot\mathbf{p}}d\mathbf{y}  \,, \nonumber
\end{eqnarray}
where $\widehat{F}\in HS(L^{2}(\mathbb{R}^{n}))$. Following the conventional terminology adopted within harmonic analysis 
\cite{Reed,Folland}, we will say that Weyl's inversion formula (\ref{InverseWeyl}) defines a {\it symbol} $f$ from its quantization $\widehat{F}$. Even though it is possible to generalize the Weyl's quantization to the space of tempered distributions $\mathcal{S}'(\mathbb{R}^{2n})$, that is, the space of continuous functionals acting on rapidly decreasing functions, also known as the Schwartz space $\mathcal{S}(\mathbb{R}^{2n})$, nevertheless, for the sake of simplicity, here we restrict our attention to integrable functions. In order to address such a generalization, we refer the reader to \cite{Folland} for a detailed exposition.  
   
\noindent With the Weyl's quantization map and its inverse at hand, we are able to obtain the Wigner function corresponding to the density operator acting on the Hilbert space $L^{2}(\mathbb{R}^{n})$. Let $\widehat{\rho}$ be the density operator associated with the quantum state $\psi\in L^{2}(\mathbb{R}^{n})$, that is, a self-adjoint, positive semi-definite operator with trace one, written as an integral operator as follows
\begin{equation}
\widehat{\rho}\varphi(\mathbf{x})=\psi(\mathbf{x})\int_{\mathbb{R}^{n}}\overline{\psi}(\mathbf{y})\varphi(\mathbf{y})d\mathbf{y}   \,,
\end{equation}
(or, equivalently, as $\widehat{\rho}=\ket{\psi}\bra{\psi}$ in Dirac notation) where $\varphi$ and 
$\psi$ belong to $L^{2}(\mathbb{R}^{n})$. By making use of the Weyl's inversion formula (\ref{InverseWeyl}), one may see that its corresponding symbol reads
\begin{equation}\label{Wigner}
\rho(\mathbf{x},\mathbf{p})=\frac{1}{(2\pi\hbar)^{n}}\int_{\mathbb{R}^{n}}\psi\left( \mathbf{x}+\frac{\mathbf{y}}{2}\right) \overline{\psi}\left( \mathbf{x}-\frac{\mathbf{y}}{2}\right) e^{-\frac{i}{\hbar}\mathbf{y}\cdot\mathbf{p}}d\mathbf{y}  \,.
\end{equation}
This special representation of the density operator stands for the celebrated Wigner function \cite{Wigner}, which allows us to characterize quantum states by using probability distributions in phase space, in a similar way as it is ordinarily formulated in statistical mechanics. However, in order to capture the probabilistic nature of quantum theory, the Wigner distribution may acquire negative values on certain regions of phase space and, therefore, this condition means that it can not be interpreted as a true probability density, and thus it is frequently referred to as a quasi-probability distribution in the literature. This apparently odd feature concerning the presence of negative values for the Wigner function, actually comprises a valuable tool which serves to characterize joint-correlation functions and entanglement properties, which in turn stand for genuine quantum attributes of mechanical systems \cite{Generating}. In addition, the Wigner function can also be used to determine the expectation values of operators by integrating their associated symbols over the phase space, namely,
\begin{equation}\label{expectation}
\braket{\psi,\widehat{A}\psi}=\int_{\mathbb{R}^{2n}}\rho(\mathbf{x},\mathbf{p})A(\mathbf{x},\mathbf{p})d\mathbf{x}d\mathbf{p}  \,,
\end{equation}   
where the symbol, $A=\mathcal{Q}^{-1}_{W}(\widehat{A})$, is obtained by applying the inverse of the Weyl's quantization map to the operator $\widehat{A}\in HS(L^{2}(\mathbb{R}^{n}))$.
Among the different remarkable properties that the Wigner function fulfills, one may emphasize that  the 
Wigner function is normalized to 1, and also that their marginal probability distributions, obtained by integrating over one of the phase space variables, prove to be linked to amplitudes either in the position or the momentum representation. These properties evidence the classical-quantum transition where solutions associated with the classical Liouville equation can be recovered in the $\hbar\to 0$ limit \cite{Decoherence}.

\noindent We turn now to the definition of the star product. Since the Weyl's map $Q_{W}:L^{2}(\mathbb{R}^{2n})\to HS(L^{2}(\mathbb{R}^{n}))$ is bijective and the product of Hilbert-Schmidt operators is closed \cite{Folland}, it implies that there is a unique function on $L^{2}(\mathbb{R}^{2n})$, denoted by $f\star g$, such that    
\begin{equation}
Q_{W}(f)Q_{W}(g)=Q_{W}(f\star g)  \,.
\end{equation}
The function $f\star g$, also called the Moyal star product between the functions
$f,g\in L^{2}(\mathbb{R}^{2n})$, is thus characterized by using the inverse Weyl's map as 
\begin{equation}\label{starWW}
(f\star g)(\mathbf{x},\mathbf{p})=Q_{W}^{-1}(Q_{W}(f)Q_{W}(g))=\tr\bigg(\widehat{\Omega}(\mathbf{x},\mathbf{p})\widehat{F}\widehat{G}\bigg)  \,.
\end{equation}
Note that, if we employ expressions (\ref{Weylmap}) and (\ref{InverseWeyl}), and after some simplifications in the exponent \cite{Cosmas}, the Moyal star product reads 
\begin{equation}\label{integralstar}
\hspace{-12ex}
(f\star g)(\mathbf{x},\mathbf{p})=\frac{1}{(\pi\hbar)^{2}}\int_{\mathbb{R}^{4}}f(\mathbf{a},\mathbf{b})g(\mathbf{c},\mathbf{d})e^{-\frac{2i}{\hbar}(\mathbf{p}\cdot(\mathbf{a}-\mathbf{c})+\mathbf{x}\cdot(\mathbf{d}-\mathbf{b})+(\mathbf{c}\cdot\mathbf{b}-\mathbf{a}\cdot{d}))}d\mathbf{a}\,d\mathbf{b}\,d\mathbf{c}\,d\mathbf{d} \,,
\end{equation}  
which is known as the integral representation of the star product and, as we can straightforwardly observe, it  ensues associativity, $(f\star(g\star h))(\mathbf{x},\mathbf{p})=((f\star g)\star h)(\mathbf{x},\mathbf{p})$, and 
also noncommutativity as a consequence of the Baker-Campbell-Hausdorff identity, as discussed in detail in \cite{Cosmas}. Further, by considering functions $f,g \in C^{\infty}(\mathbb{R}^{2n})$, one may Taylor expand both functions inside the integral (\ref{integralstar}), thus yielding the Moyal star product in the differential representation \cite{Teaching},
\begin{eqnarray}\label{Star}
(f\star g)(\mathbf{x},\mathbf{p})&=&\sum_{m,n=0}^{\infty}\left( \frac{i\hbar}{2}\right)^{m+n}\frac{(-1)^{m}}{m!n!}\left( \partial^{m}_{\mathbf{p}}\cdot\partial^{n}_{\mathbf{x}}f\right)\left( \partial^{n}_{\mathbf{p}}\cdot\partial^{m}_{\mathbf{x}}g\right) \nonumber \\  
&=&f(\mathbf{x},\mathbf{p})\exp\left[\frac{i\hbar}{2}\left(\frac{\overleftarrow{\partial}}{\partial \mathbf{x}}\cdot\frac{\overrightarrow{\partial}}{\partial\mathbf{p}}- \frac{\overleftarrow{\partial}}{\partial \mathbf{p}}\cdot\frac{\overrightarrow{\partial}}{\partial\mathbf{x}}\right)\right]g(\mathbf{x},\mathbf{p})   \,,  
\end{eqnarray}   
where the differential operators $\overleftarrow{\partial}/\partial \mathbf{x}$, $\overleftarrow{\partial}/\partial \mathbf{p}$ act on the left, while the differential operators $\overrightarrow{\partial}/\partial \mathbf{x}$, $\overrightarrow{\partial}/\partial \mathbf{p}$ act on the right. As pointed out in \cite{Bayen}, the star product defined on equation (\ref{Star}),  comprises an associative formal deformation of $C^\infty(\mathbb{R}^{2n})$ given by
\begin{equation}
f\star g=fg+\sum_{n=1}^\infty \left(\frac{i\hbar}{2} \right)^{n} P^{n}(f,g) \,,
\end{equation}  
where the $\mathbb{R}$-bilineal map $P^{n}:C^{\infty}(\mathbb{R}^{2n})\times C^\infty(\mathbb{R}^{2n})$ corresponds to the $n$th power of the Poisson bracket, interpreted as a bidifferential operator acting on $f$ and $g$,
\begin{equation}
\left\lbrace f,g\right\rbrace=\frac{\partial f}{\partial \mathbf{x}}\cdot\frac{\partial g}{\partial \mathbf{p}}-\frac{\partial f}{\partial \mathbf{p}}\cdot\frac{\partial g}{\partial \mathbf{x}}=P(f,g)  \,.
\end{equation}
Moreover, by defining the Moyal bracket as
\begin{equation}
\left\lbrace f,g \right\rbrace_{M}=\frac{1}{i\hbar}\left(f\star g-g\star f \right)  \,,  
\end{equation}
we may note that it is this bracket, and not the Poisson bracket, the one that corresponds to the quantum commutator under the Weyl's quantization map, namely
\begin{equation}
\left\lbrace f,g \right\rbrace_{M}=\frac{1}{i\hbar}\mathcal{Q}^{-1}_{W}\bigg(\left[\mathcal{Q}_{W}(f),\mathcal{Q}_{W}(g) \right]  \bigg)  \,,
\end{equation}
and its expansion in terms of the parameter $\hbar$ indicates that the Moyal bracket consists of a quantum correction of the Poisson bracket.
Analogously to the standard Hilbert space conventionalism, within the deformation quantization approach the spectral properties of operators are provided by star-genvalues equations. For the case of a stationary Wigner function (\ref{Wigner}), and given that star product involves exponentials of derivative operators, we have  
\begin{equation}
H(\mathbf{x},\mathbf{p})\,\star \, \rho(\mathbf{x},\mathbf{p})=H\left(\mathbf{x}+\frac{i\hbar}{2}\frac{\overrightarrow{\partial}}{\partial\mathbf{p}}, \mathbf{p}- \frac{i\hbar}{2}\frac{\overrightarrow{\partial}}{\partial\mathbf{x}}\right)\rho(\mathbf{x},\mathbf{p})=E\rho(\mathbf{x},\mathbf{p})  \,, 
\end{equation}   
where $H(\mathbf{x},\mathbf{p})$ is the classical Hamiltonian function and $E$ corresponds to the energy eigenvalues of the quantum Hamiltonian operator, $\widehat{H}\psi=E\psi$, in the Schr\"odinger representation \cite{Curtright}.  

Let $\widehat{\rho}$ be a density operator taken in the Schr\"odinger picture, and $\widehat{H}$ the Hamiltonian operator, the time evolution of $\widehat{\rho}$ is described by the von Neumann equation \cite{Cohen},
\begin{equation}
\frac{\partial\widehat{\rho}}{\partial t}=-\frac{i}{\hbar}\left[ \widehat{H},\widehat{\rho}\right]  \,. 
\end{equation} 
In particular, if the Hamiltonian operator is time independent, the von Neumann equation can be solved as
\begin{equation}
\widehat{\rho}(t)=\widehat{U}(t)\widehat{\rho}(0)\widehat{U}^{\dagger}(t)  \,,
\end{equation}
where $\widehat{U}(t)=e^{-i\widehat{H}t/\hbar}$ corresponds to a strongly continuous one parameter group of unitary operators associated with the self-adjoint operator $\widehat{H}$ 
which, in consequence, generates the time evolution in quantum dynamics \cite{Takhtajan}.
Within the deformation quantization approach, the von Neuman equation is translated, by means of the Weyl's correspondence, into the Moyal's equation of motion for the Wigner function, namely, 
\begin{equation}\label{Moyal}
\frac{\partial \rho(\mathbf{x},\mathbf{p})}{\partial t}=-\frac{i}{\hbar}\left( H\star\rho-\rho\star H\right)=\left\lbrace \rho,H \right\rbrace_{M}=P(\rho,H)+\mathcal{O}(\hbar^2)  \,.  
\end{equation}
This is an extension of Liouville's theorem in classical mechanics, since as $\hbar\to 0$, one recovers to some extent the classical behaviour of the system.
The solution of the Moyal's dynamical equation (\ref{Moyal}) is obtained formally in terms of the star exponential \cite{Bayen},
\begin{equation}\label{Exp}
\Exp\left(-\frac{i}{\hbar}tH \right)\equiv \sum_{n=0}^{\infty}\frac{1}{n!}\left(-\frac{i}{\hbar}t \right)^{n}H^{*n}   \,,  
\end{equation}  	
with $H^{*n}=H\star H\cdots\star H$ ($n$ factors) as,
\begin{equation}
\rho(t)=\Exp\left(-\frac{i}{\hbar}tH \right)\star \rho\star \Exp\left(\frac{i}{\hbar}tH \right)   \,,
\end{equation}
where $\rho(t)$ stands for the time evolved Wigner function at time $t$. The star exponential (\ref{Exp}) can be considered as an element of the space of formal series in $\hbar$ with coefficients in $C^{\infty}(\Gamma)$, usually denoted by $C^{\infty}(\Gamma)[[\hbar]]$, which for many relevant physical examples, converges in the distributional sense \cite{damped}. The star exponential, considered as a distribution, holds a Fourier-Dirichlet expansion \cite{Bondia},
\begin{equation}\label{FDexpansion}
\Exp\left(-\frac{i}{\hbar}tH \right)=\sum_{n=0}^{\infty}e^{-\frac{i}{\hbar}tE_{n}}\rho_{n}  \,,
\end{equation} 
where $E_{n}$ denotes the eigenvalues of the Hamiltonian operator $\widehat{H}$, and the functions $\rho_{n}$ represent the symbols obtained by applying the inverse of Weyl's quantization map (\ref{InverseWeyl}) to the projection operator $\widehat{P}_{\ket{n}}=\ket{n}\bra{n}$, associated with the normalized energy eigenstate $\ket{n}$, and such that
\begin{eqnarray}
H\star\rho_{n} & = & E_{n}\rho_{n}   \,, \nn\\ 
\rho_{m}\star\rho_{n} & = & \delta_{mn}\rho_{n}  \,.
\end{eqnarray}
Moreover, for \blue{a} continuous spectrum the summations in the former relations must be extended to integrals over a continuous parameter, given by the energy, and the Kronecker delta functions are replaced by Dirac delta distributions.
  
Given a Hamiltonian $H(\mathbf{x},\mathbf{p})$, in order to determine the star exponential function (\ref{Exp}), we note that
\begin{equation}
\label{eq:HStarExp}
H\star\Exp\left(-\frac{i}{\hbar}tH \right)=i\hbar\frac{d}{dt}\Exp\left(-\frac{i}{\hbar}tH \right)  \,,
\end{equation}
then, as was proven in \cite{Bayen}, a solution of the former differential equation, denoted by $\phi(t,H)$, can be constructed only if $\phi(t,H)$, written as a power series of $t$, converges on $\mathcal{D}'(\mathbb{R}^{2n})$ in the sense of distributions, i.e., $\phi(t,H)$ stands for a continuous, linear functional acting on smooth functions with compact support $C^{\infty}_{c}(\mathbb{R}^{2n})$. However, in the vast majority of the physically relevant cases, investigation of convergence  of star products may 
confront major obstacles in view of the highly nontrivial analytic structure of the formal series appearing in~(\ref{FDexpansion}) and~(\ref{eq:HStarExp}). In addition, as mentioned on \cite{Waldmann}, the quest for convergence remains as one of the most important open problems in deformation quantization. Furthermore, the star exponential functions not only enables us to determine the time evolution of the Wigner function and other quantum observables but it also allows us to quantize gauge systems by constructing both a physical Wigner function and a physical inner product by implementing a star group averaging on the constraints consisting in integrating the star exponential of the classical constraints functions with respect to some appropriate gauge invariant measure \cite{SGA}. In order to overcome some of the above mentioned difficulties, in the next section we will investigate the properties of the star exponential in connection with the path integral formalism.          
 
\section{Star exponentials via quantum propagators}  
\label{sec:StarExp}

By means of the inverse Weyl's quantization map  previously analyzed, in this section we will determine the star exponential of the Hamiltonian function in terms of quantum mechanical propagators and path integrals. For simplicity, we will limit our attention to systems with one degree of freedom, but the generalization to more dimensions follows straightforwardly. We conclude this section illustrating the proposed approach with a couple of relevant physical examples. 

\subsection{Propagators and star exponentials}
We will start by considering that the Hamiltonian of a quantum particle moving under the influence of a potential $V(x)$ is given by
\begin{equation}
\widehat{H}=\frac{\widehat{P}^{2}}{2m}+V(\widehat{X})    \,,
\end{equation}
then, the Cauchy initial value problem in coordinate representation reads
\begin{equation}\label{Schrodinger}
i\hbar\frac{\partial\psi}{\partial t}=-\frac{\hbar^{2}}{2m}\frac{\partial^{2}\psi}{\partial x^{2}}+V(x)\psi  \,,
\end{equation} 
subject to an initial condition of the form   
\begin{equation}
\psi(x,t)\big|_{t=t_{0}}=\psi(x_0,t_{0})  \,.
\end{equation} 
Under reasonably general conditions on the potential $V(x)$, such as local integrability and boundedness, the former Cauchy problem presents a fundamental solution, denoted by $K(x,t,x_{0},t_{0})$, which satisfies the Schr\"odinger equation (\ref{Schrodinger}) in the weak distributional sense \cite{Reed}, and the initial condition
\begin{equation}
\lim_{t\to t_{0}}K(x,t,x_{0},t_{0})=\delta(x-x_{0})   \,.
\end{equation}   
Consequently, the solution of the Cauchy problem can be written as
\begin{equation}
\psi(x,t)=\int_{\mathbb{R}}K(x,t,x_{0},t_{0})\psi(x_{0},t_{0})dx_{0}   \,,
\end{equation} 
where the integration should be also understood in the distributional sense. The value of the function $|K(x,t,x_{0},t_{0})|^{2}$ represents the conditional probability distribution of finding a quantum particle at the point $x\in\mathbb{R}$ at time $t$, assuming that it was at the point $x_{0}\in\mathbb{R}$ at time $t_{0}$. In physics, the fundamental solution $K(x,t,x_{0},t_{0})$ corresponds to the distributional kernel of the evolution operator $\widehat{U}(t-t_{0})$ and is called the quantum mechanical propagator or simply the amplitude. In Dirac's terminology, the propagator can also be written as
\begin{equation}
K(x,t,x_{0},t_{0})=\braket{x,t|x_{0},t_{0}}=\braket{x|e^{-\frac{i}{\hbar}\widehat{H}(t-t_{0})}|x_{0}}   \,.
\end{equation} 
Assuming the spectrum of the Hamiltonian operator is known, it is possible to express the propagator in a closed form
\begin{equation}
K(x,t,x_{0},t_{0})=\sum_{n=0}^{\infty}e^{-\frac{i}{\hbar}E_{n}(t-t_{0})}\psi_{n}(x)\overline{\psi}_{n}(x_{0})    \,,
\end{equation}  
where the functions, $\left\lbrace\psi_{n}(x) \right\rbrace_{n=0}^{\infty}\in L^{2}(\mathbb{R})$, comprise a set of eigenfunctions of $\widehat{H}$ with eigenvalues $E_{n}$ and the series converges in a distributional sense, providing a representation of the propagator in terms of the spectral decomposition of the Hamiltonian operator \cite{Takhtajan}. 

Our next task is to express the quantum propagator $K(x_{f},t_{f},x_{0},t_{0})$, as an integral transform of the star exponential via the symbol associated to the evolution operator through the inverse Weyl's quantization map. Since the star product obtained in (\ref{starWW}) defines a homomorphism between the classical observables, $C^ {\infty}(\mathbb{R}^{2})$, and linear operators acting on the Hilbert space $L^{2}(\mathbb{R})$, this implies that the symbol corresponding to the formal evolution operator 
\begin{equation}
e^{-\frac{i}{\hbar}t\widehat{H}}=1-\frac{it}{\hbar}\widehat{H}+\frac{1}{2!}\left( \frac{-it}{\hbar}\right)^{2}\widehat{H}^{2}+\cdots   \,, 
\end{equation} 
is given by the star exponential \cite{Bayen}, \cite{Curtright},
\begin{equation}
\Exp{\left( -\frac{i}{\hbar}tH\right)}=1-\frac{it}{\hbar}H+\frac{1}{2!}\left( \frac{-it}{\hbar}\right)^{2}H\star H+\cdots   \,.
\end{equation}  
Let $\widehat{\rho}_{f,0}=\ket{x_{0}}\bra{x_{f}}$ be a nondiagonal density operator, with $\ket{x_{0}}$ and $\ket{x_{f}}$ being eigenvectors associated to the position operator $\widehat{X}$, such that, $\widehat{X}\ket{x_{0}}=x_{0}\ket{x_{0}}$ and $\widehat{X}\ket{x_{f}}=x_{f}\ket{x_{f}}$, respectively, where $x_{0},x_{f}\in\mathbb{R}$. Using the inverse Weyl's inversion formula (\ref{InverseWeyl}), the corresponding nondiagonal Wigner function related to the density operator $\widehat{\rho}_{f,0}$ reads
\begin{equation}
\rho_{f,0}(x,p) := \mathcal{Q}^{-1}_{W}(\ket{x_{0}}\bra{x_{f}})=\frac{1}{2\pi\hbar}e^{\frac{i}{\hbar}(x_{f}-x_{0})p}\delta\left( x-(x_{f}+x_{0})/2\right). 
\end{equation}    
Using the integral property of the Wigner function (\ref{expectation}) in the sense of distributions, the propagator may be identified as follows  
\begin{eqnarray}\label{Kstar}
\hspace{-11ex}
K(x_{f},t_{f},x_{0},t_{0})&=&\int_{\mathbb{R}^{2}}\rho_{f,0}(x,p)\, \Exp{\left(-\frac{i}{\hbar}(t_f - t_0)H(x,p) \right)} \, dxdp \nn\\
&=&\frac{1}{2\pi\hbar}\int_{\mathbb{R}}e^{\frac{i}{\hbar}(x_{f}-x_{0})p}\, \Exp{\left( -\frac{i}{\hbar}(t_{f}-t_{0})H\left((x_{f}+x_{0})/2,p \right)\right)}\,dp  \,. 
\end{eqnarray} 
The former expression shows that the propagator can be formulated as the Fourier transform of a star exponential of the Hamiltonian function defined on the classical phase space \cite{Sharan}. To proceed further, we introduce the new variables $q=(x_{f}+x_{0})/2$, $q'=(x_{f}-x_{0})/2$ and we set, for the sake of simplicity, $t_{f}=t$ and $t_{0}=0$. By computing the inverse Fourier transform of the formula (\ref{Kstar}), we have
\begin{equation}\label{StarE}
\Exp\left(-\frac{i}{\hbar}tH(q,p)\right)=2\int_{\mathbb{R}}e^{-\frac{i}{\hbar}2q'p}K(q+q',t,q-q',0)\, dq' \,.
\end{equation}  
Equation (\ref{StarE}) provides us with an advantageous approach to calculate star exponentials 
without relying on convergence of formal series, since as mentioned in the previous section, this can become rather difficult. Instead, it is possible to make use of the Feynman's representation of the propagator in terms of path integrals, i.e., express the propagator, $K(q+q',t,q-q',0)$, as a weighted sum over all possible histories of the classical motion in the phase space since, as we will observe in the following subsection, certain aspects of quantum mechanics become more transparent within this formulation such as for instance, the fact that   probability amplitudes of quadratic Lagrangians involve the classical action, as well as the convenience of using this framework in order to generalize to the case of field theories \cite{Takhtajan,Baaquie}. 

\subsection{Examples}
In this subsection we present some well-known examples with the main aim to explicitly illustrate how the star exponential can be obtained via propagators and path integrals through the method developed above. 
\subsubsection{Free particle.}
As a first test case, let us consider the quantum propagator for a free particle in one dimension \cite{Takhtajan},
\begin{equation}
K_{\mathrm{free}}(x_{f},t_{f},x_{0},t_{0})=\left(\frac{m}{2\pi i \hbar(t_{f}-t_{0})}\right)^{1/2}\exp\left[ \frac{im}{2\hbar(t_{f}-t_{0})}(x_{f}-x_{0})^{2}\right]  \, .  
\end{equation}
According to (\ref{StarE}), the star exponential corresponding to the free particle Hamiltonian, $H_{\mathrm{free}}(q,p)=p^{2}/2m$, reads  
\begin{eqnarray}
\Exp\left(-\frac{i}{\hbar}tH_{\mathrm{free}}(q,p)\right)&=& 2\int_{\mathbb{R}}e^{-\frac{i}{\hbar}2q'p}K_{\mathrm{free}}(q+q',t,q-q',0)dq' \nonumber \\
&=& e^{-\frac{it}{\hbar}\left(\frac{p^{2}}{2m}\right)} \, .
\end{eqnarray}
In general, since the star exponential of any Hamiltonian function can be computed by means of its spectrum (\ref{FDexpansion}), in case of a continuous spectrum we have
\begin{equation}\label{SEspect}
\Exp{\left(-\frac{i}{\hbar}H(q,p)\right)}=\int_{R}e^{-\frac{i}{\hbar}tE}\rho_{E}(q,p) \ dE, 
\end{equation}  
where $\rho_{E}(q,p)$ corresponds to the diagonal Wigner distribution associated with an eigenstate of the Hamiltonian with energy $E$, as described in detail in \cite{Generating}. Once the star exponential has been computed, we can make use of the Fourier inversion formula on (\ref{SEspect}) to obtain the corresponding Wigner distribution for a free particle as follows
\begin{eqnarray}
\rho_{E}(q,p)&=&\frac{1}{2\pi\hbar^{2}}\int_{\mathbb{R}}e^{-\frac{i}{\hbar}tH_{\mathrm{free}}(q,p)}e^{\frac{i}{\hbar}tE}  \, dt \nonumber\\
&=&\frac{1}{2\pi\hbar}\delta(p^{2}/2m-E)   \, ,
\end{eqnarray}
which coincides with the integral representation of the Wigner distribution, derived directly from the wave function originated from the Schr\"odinger equation \cite{Curtright}. 

\subsubsection{Harmonic oscillator.}
Our second example is devoted to analyze the case of a quantum harmonic oscillator in one dimension, with Hamiltonian $H_{\mathrm{ho}}(q,p)=p^{2}/2m+m\omega^{2}q^{2}/2$. For this instance, the quantum propagator can be written as \cite{Takhtajan},
\begin{equation}
\hspace{-13ex}
K_{\mathrm{ho}}(x_{f},t_{f},x_{0},t_{0})=\sqrt{\frac{m\omega}{2\pi i\hbar\sin(\omega T)}}\exp\left[\frac{im\omega}{2\hbar\sin(\omega T)}\left( (x_{f}^{2}+x_{0}^{2})\cos(\omega T)-2x_{f}x_{0}\right) \right]   
\end{equation} 
where the parameter $T$ is given by $T=t_{f}-t_{0}\in\mathbb{R}$. From (\ref{StarE}) we can read off that
\begin{eqnarray}\label{SEho}
\Exp\left(-\frac{i}{\hbar}tH_{\mathrm{ho}}(q,p)\right)&=& 2\int_{\mathbb{R}}e^{-\frac{i}{\hbar}2q'p}K_{\mathrm{ho}}(q+q',t,q-q',0)dq' \nonumber \\
&=& \left(\cos\left( \frac{\omega t}{2}\right)  \right)^{-1}\exp\left[\frac{2H_{\mathrm{ho}}(q,p)}{i\omega\hbar}\tan\left(\frac{\omega t}{2} \right)  \right]  \,.  
\end{eqnarray}
For a fixed $t\in\mathbb{C}$, with $\Imp{t}\leq 0$ and $t\neq (2k+1)\pi$, having $k\in \mathbb{Z}$, the preceding star exponential owns a Fourier-Dirichlet expansion given by (see reference \cite{Bayen} for details),
\begin{equation}
 \left(\cos\left( \frac{\omega t}{2}\right)  \right)^{-1}\exp\left[\frac{2H_{\mathrm{ho}}(q,p)}{i\omega\hbar}\tan\left(\frac{\omega t}{2} \right)  \right]=\sum_{n=0}^{\infty}\rho_{n}(q,p)e^{-it\omega(n+1/2)}  \,,
\end{equation}
where the diagonal Wigner functions, $\rho_{n}(q,p)$, associated with the energy eigenvectors with eigenvalues $E_{n}=\hbar\omega(n+1/2)$ are
\begin{equation}
\rho_{n}(q,p)=\frac{(-1)^{n}}{\pi\hbar}e^{-\frac{2H_{\mathrm{ho}}(q,p)}{\omega\hbar}}L_{n}\left(\frac{4H(q,p)}{\omega\hbar} \right)   \,.
\end{equation}
Here $L_{n}\left(\frac{4H(q,p)}{\omega\hbar} \right)$ denotes the usual Laguerre polynomials of order $n$. It is worth mentioning that the expression for the star exponential (\ref{SEho}), derived from the propagator, agrees with those obtained by employing star eigenvalue equations, but without relying on convergence issues of formal series and partial differential equations \cite{Curtright}, which in certain scenarios, can be rather challenging. 

\subsubsection{Linear potential.}
Let us consider now the Hamiltonian $H_{\mathrm{lp}}(q,p)=p^{2}+q$, which corresponds to a linear potential in one dimension. In this case the quantum propagator is given by \cite{Baaquie},
\begin{equation}
\hspace{-5ex}
K_{\mathrm{lp}}(x_{f},t_{f},x_{0},0)=\sqrt{\frac{1}{4\pi i\hbar t}}\exp\left[\frac{i}{4t\hbar}(x_{f}-x_{0})^{2}+\frac{it}{2\hbar}(x_{f}+x_{0})-\frac{it^{3}}{12\hbar} \right]  \,. 
\end{equation}     
Based on the expression (\ref{StarE}), the star exponential for the linear potential follows
\begin{eqnarray}
\Exp\left(-\frac{i}{\hbar}tH_{\mathrm{lp}}(q,p)\right)&=& 2\int_{\mathbb{R}}e^{-\frac{i}{\hbar}2q'p}K_{\mathrm{lp}}(q+q',t,q-q',0) \, dq' \nonumber \\
&=& e^{-\frac{i}{\hbar}t(q+p^{2}+t^{2}/12)}  \,,
\end{eqnarray}
which coincides with the combinatoric derivation through generating functions obtained in \cite{Generating}. Then, since this model admits a continuous spectrum, the Wigner function can  be determined by means of the Fourier inversion formula in expression (\ref{SEspect}), thus obtaining
\begin{eqnarray}
\rho_{E}(q,p)&=&\frac{1}{2\pi\hbar^{2}}\int_{\mathbb{R}}e^{-\frac{i}{\hbar}t(q+p^{2}+t^{2}/12)}e^{\frac{i}{\hbar}tE}   \, dt \nonumber\\
&=& \frac{2^{2/3}}{2\pi\hbar}\Ai{\left(\frac{2^{2/3}}{\hbar}(p^{2}+q-E) \right)}  \,,
\end{eqnarray} 
where $\Ai(x)$ denotes the Airy functions. The former result proves to be in accordance with the Wigner function computed in \cite{Generating}, in terms of the eigenfunctions of the Hamiltonian operator.

\subsubsection{General quadratic Lagrangian.} 
Our next example is devoted to theories defined by a general one-dimensional quadratic Lagrangian of the form
\begin{equation}\label{Lq}
L(q(t),\dot{q}(t),t)=\frac{m}{2}\dot{q}^{2}-\frac{c(t)}{2}q^{2}+f(t)q  \,,
\end{equation}   
where $c(t), f(t)\in C(\mathbb{R})$. From the path integral formulation of quantum mechanics \cite{Baaquie, Zinn}, the propagator can be written as a path integral in position space as follows
\begin{equation}\label{path}
K(q_{f},t_{f},q_{0},t_{0})=\int\mathcal{D}q(t)\, e^{\frac{i}{\hbar}S(q(t))}  \,,
\end{equation}
where the integration measure $\mathcal{D}q(t)$ can be regarded as the limit
\begin{equation}
\mathcal{D}q(t)=\lim_{N\to\infty}\left(\frac{m}{2\pi i\hbar \Delta t} \right)^{\frac{N+1}{2}}\int_{\mathbb{R}}\prod_{k=1}^{N}dq_{k}   \,, 
\end{equation}
for $\Delta t=(t_{f}-t_{0})/(N+1)$ stands for a partition of the time interval $\left[t_0,t_f \right]$, while $S(q(t))$ denotes the classical action
\begin{equation}
S(q(t))=\int_{t_{0}}^{t_{f}}L(q(t),\dot{q}(t),t)\,   dt  \,.
\end{equation}
Formula (\ref{path}) needs to be handled with caution as it can be understood as an integral over all classical paths satisfying the boundary conditions 
\begin{equation}
q(t_{0})=q_{0} \,, \;\;\; q(t_{f})=q_{f}   \,,
\end{equation}  
which, certainly, should not be interpreted as an integral in the sense of measure theory, since it is a well-known fact that an infinite dimensional vector space does not possess a measure playing the role of a Lebesgue measure \cite{Takhtajan}. Bearing this in mind, let us denote by $q_{c}(t)$ the solution to the Euler-Lagrange equations of motion fulfilling the boundary conditions $q_{c}(t_{0})=q_{0}$ and $q_{c}(t_{f})=q_{f}$.   
Within the Feynman path integral formulation, the quantum propagator takes the form \cite{Zinn},
\begin{equation}
K(q_{f},t_{f},q_{0},t_{0})=\sqrt{\frac{m}{2\pi i\hbar \phi(t_{f})}}e^{\frac{i}{\hbar}S_{c}(q_{f},t_{f},q_{0},t_{0})}   \,,
\end{equation}
where $S_{c}(q_{f},t_{f},q_{0},t_{0})$ denotes the action evaluated at $q_{c}(t)$ and the function, $\phi(t)\in C^{2}(\mathbb{R})$, arises from analyzing the behavior of the classical momentum variable under the discretization related to the path integral measure \cite{Zinn}, which satisfies the differential equation
\begin{equation}
m\frac{d^{2}\phi}{dt^{2}}+c(t)\phi(t)=0  \,,
\end{equation}    
subject to the boundary conditions $\phi(t_{0})=0$ and $\dot{\phi}(t_{0})=1$, and
\begin{equation}
\frac{1}{\sqrt{\phi(t_{f})}}=e^{-i\pi\nu/2}\frac{1}{\sqrt{|\phi(t_{f})|}},
\end{equation}
where the parameter $\nu$ is called the Maslov index \cite{Maslov} and it depends on the values $t_0$ and $t_{f}$. Subsequently, the star exponential   corresponding to the Hamiltonian function $H(q,p)$, associated with the general one dimensional quadratic Lagrangian (\ref{Lq}) can be expressed as follows
\begin{eqnarray}\label{TO}
\Exp\left(-\frac{i}{\hbar}tH(q,p)\right)&=& 2\int_{\mathbb{R}}e^{-\frac{i}{\hbar}2q'p}K(q+q',t,q-q',0)  \, dq' \nonumber \\
&=& \sqrt{\frac{2m}{\pi i\hbar \phi(t_{f})}}\int_{\mathbb{R}}e^{-\frac{i}{\hbar}2q'p} e^{\frac{i}{\hbar}S_{c}(q+q',t,q-q',0)}  \, dq'   \,.
\end{eqnarray}
Properly speaking, since the Lagrangian defined on (\ref{Lq}) is time dependent, the star exponential (\ref{TO}) corresponds to the symbol associated to the time ordered evolution operator, usually written in terms of Dyson's series \cite{Takhtajan}. Therefore, we notice that the resulting description not only reproduce the familiar results but it also may extend the applicability of the deformation quantization program to more involved and interesting physical systems.

\subsubsection{Particle constrained to a circle.}
As our last example we consider a particle constrained to move on a circle $S^{1}$ parametrized with the angle $\varphi$, such that $0\leq \varphi\leq 2\pi$, with $\varphi=0$ and $\varphi=2\pi$ identified, this means that the model is subject to periodic boundary conditions. The Lagrangian of the system is given by $L_{\mathrm{c}}=I\dot{\varphi}^{2}$, where $I$ stands for the moment of inertia. For this example the propagator reads \cite{Schulman},       
\begin{equation}
\hspace{-5ex}
K_{\mathrm{c}}(\varphi_{f},t,\varphi_{i},0)=\left( \frac{I}{2\pi i\hbar t}\right)^{1/2}\exp\left[\frac{i}{\hbar}\frac{I}{2t}(\varphi_{f}-\varphi_{i})\right]\theta_{3}\left(\frac{\pi I(\varphi_{f}-\varphi_{i})}{\hbar t}\bigg|\frac{2\pi I} {\hbar t} \right),  
\end{equation}
where $\theta_{3}$ is given by the Jacobi's theta function \cite{Schulman,Grad},
\begin{equation}\label{Theta}
\theta_{3}(z|t)=\sum_{n=-\infty}^{\infty}e^{i\pi t n^{2}+i 2 n z}=\theta_{3}(-z|t).
\end{equation} 
By employing the infinite sum representation of the Jacobi's theta function (\ref{Theta}) and its quasiperiodic properties $\theta_{3}(z+\pi|t)=\theta_{3}(z|t)$ and $\theta_{3}(z+\pi t|t)=e^{-i\pi t-2iz}\theta_{3}(z|t)$, the propagator can be written as
\begin{equation}
K_{\mathrm{c}}(\varphi_{f},t,\varphi_{i},0)=\frac{1}{2\pi}\sum_{n=-\infty}^{\infty}\exp\left[\frac{-i\hbar n^{2}t}{2I} \right]e^{i n(\varphi_{f}-\varphi_{i})}.   
\end{equation}
By substituting the former expression in the formula (\ref{StarE}), the star exponential for a particle moving on a circular orbit, we obtain
\begin{equation}\label{SEc}
\Exp\left(-\frac{i}{\hbar}tH_{\mathrm{c}}(q,p)\right)=\sum_{n=-\infty}^{\infty}e^{-\frac{i}{\hbar}\frac{\hbar^{2}n^{2}}{2I}}\frac{1}{2\pi}\sinc \left[ \pi(p-n)\right],
\end{equation}
where the $\sinc$ function is defined by
\begin{equation}
\sinc\left[\pi(p-n) \right]=\frac{\sin\,\pi(p-n)}{\pi(p-n)}, \;\; p\in\mathbb{R}, \; n\in\mathbb{Z}.
\end{equation}
The star exponential (\ref{SEc}) allows us to identify in an straightforward manner that the Wigner function for a particle moving in a closed path is given by
\begin{equation}
\rho_{n}(\varphi,p)=\frac{1}{2\pi}\sinc \left[ \pi(p-n)\right],
\end{equation}
with energy eigenvalues $E_{n}=\hbar^{2}n^{2}/2I$. The preceding Wigner function $\rho_{n}(\varphi,p)$ coincides with the Wigner function computed in \cite{Kastrup}, where the Wigner operator associated to the cylindrical phase space is constructed by means of unitary representations of the Euclidean group.

\subsection{Path integral representation of the star product}

Finally, let us conclude this section by representing the star product, introduced in (\ref{integralstar}), in terms of path integrals. We are particularly interested in computing the correlation functions between two arbitrary operators $\widehat{F}$ and $\widehat{G}$ in the Heisenberg picture, as
\begin{equation}
\hspace{-14ex}
\braket{q_{f},t_{f}|\widehat{F}(\widehat{q}(t_{2}),\widehat{p}(t_{2}))\widehat{G}(\widehat{q}(t_{1}),\widehat{p}(t_{1}))|q_{0},t_{0}}=\braket{q_{f}|e^{\frac{i}{\hbar}(t_{2}-t_{f})\widehat{H}}\widehat{F}(\widehat{q},\widehat{p})e^{\frac{i}{\hbar}(t_{1}-t_{2})\widehat{H}}\widehat{G}(\widehat{q},\widehat{p})e^{\frac{i}{\hbar}(t_{0}-t_{1})\widehat{H}}|q_{0}},
\end{equation}
where $\widehat{H}$ is the Hamiltonian operator of the quantum system and $t_{0}<t_{1}\leq t_{2}<t_{f}$. By using the Weyl's inversion formula (\ref{InverseWeyl}) on the right hand side of the former equation, we have
\begin{equation}
\hspace{-10ex}
\label{eq:correlation2}
\braket{q_{f},t_{f}|\widehat{F}(\widehat{q}(t_{2}),\widehat{p}(t_{2}))\widehat{G}(\widehat{q}(t_{1}),\widehat{p}(t_{2}))|q_{0},t_{0}}=\frac{1}{2\pi\hbar}\int_{\mathbb{R}}e^{\frac{i}{\hbar}(q_{f}-q_{0})p}B\left(\frac{q_{f}+q_{0}}{2},p\right)dp, 
\end{equation}     
where the function $B(q,p)$ is given by 
\begin{equation}
\hspace{-5ex}B(q,p)=U_{\star}(t_{2},t_{f})\star f(q,p)\star U_{\star}(t_{1},t_{2})\star g(q,p)\star U_{\star}(t_{0},t_{1})  \,,
\end{equation}
and we have defined $U_{\star}(t_{a},t_{b})$ as 
\begin{equation}
U_{\star}(t_{a},t_{b})=\Exp\left(\frac{i}{\hbar}(t_{a}-t_{b})H(q,p)\right)
\end{equation}
while $f$, $g$ correspond to the symbols associated with the operators $\widehat{F}, \widehat{G}$ respectively. To proceed further, we introduce the change of variables $x=(q_{f}+q_{0})/2$, $y=(q_{f}-q_{0})/2$. If we now use the path integral expression for the correlation functions~(\ref{eq:correlation2}), setting the Hamiltonian $\widehat{H}$ equal to zero, and performing an inverse Fourier transform, we finally obtain
\begin{equation}\label{starpath}
\hspace{-14ex}
(f\star g)(x,p)=2\int_{{\mathbb{R}}}e^{-\frac{i}{\hbar}2yp}\left(\int\mathcal{D}p(t)\mathcal{D}q(t)f(q(t_{2}),p(t_{2}))g(q(t_{1}),p({t_{1}}))e^{\frac{i}{\hbar}\int_{x-y}^{x+y}p dq }\right)dy   \,,
\end{equation}
where the points $(q(t_{1}),p({t_{1}}))$ and $(q(t_{2}),p(t_{2}))$ are being integrated over the path integral. Even though this last result corresponds to the case of a quantum system with finite degrees of freedom, one may notice that it agrees with the analogous result provided by Cattaneo and Felder in \cite{Cattaneo}, where a relation between the Deformation Quantization formalism and the quantum field theory of a Poisson Sigma model with generalized gauge fields was first elucidated.\footnote{It is known that in phase space the boundary conditions are overdetermined and, consequently,
the boundary conditions are not elliptic, thus the path integral is not well defined~\cite{Gaiotto:2021kma}. However, formula (\ref{starpath}) still is valid when the particle is moving on a circle and $t_1$ and $t_2$ belongs to the circle as, in such a case, it is not necessary to choose boundary conditions and the path integral is well-defined giving rise to a genuine Deformation Quantization.  This issue was discussed in detail for the Poisson sigma model in~\cite{Cattaneo}, \cite{Grady:2015ica}, where different boundary conditions are considered for all of the involved fields.} 
As already mentioned in \cite{Path}, equation (\ref{starpath}) ends up being formally similar to the Moyal product in the symplectic case. However, we expect that the implementation of the propagator, instead of a topological field theory, will provide us with optimal strategies to analyze the star product in more general scenarios such as curved spacetimes and quantum gravity.

\section{Conclusions}
\label{sec:conclu}

Star exponential functions are fundamental in the Deformation Quantization program as they stand for the analogous of the time evolution operator.   However, the issue of convergence for such functions is still an open mathematical problem.  Besides, very few examples are reported in the literature as, in practice, it may be a difficult task to obtain analytic expressions for the star exponential functions.  In this paper, we addressed the relation between the star exponential functions and propagators in Quantum Mechanics which, in turn, may be explicitly realized by means of the celebrated Feynman's path integral formulation.   Indeed, here we analyzed the star exponential of the classical Hamiltonian function by means of an appropriate Fourier transform of the quantum mechanical propagator providing us with a reliable way to obtain star exponentials without depending on the issue of convergence of the formal series involved in the Moyal star product formalism.  As discussed above, our developments may become more relevant by introducing the well-known Feynman's path integral representation of the propagator as a sum over all the classical histories in phase space.   In particular, we included some physically motivated examples for which one may conveniently acknowledge the construction of the star exponential functions via Feynman's path integrals, recovering results previously reported in the literature.
Within our formalism, we were also able to find an analogous expression for the star product developed by Cattaneo and Felder in the context 
of Kontsevich's quantization formula.  We expect that this result may be directly generalized to obtain the Moyal product for any symplectic manifold.

We expect that the construction of star exponential functions addressed here may pave the way not only towards the understanding of convergence issues  but also to the application of such methods for non-polynomial systems.  In particular, a compelling direction of study would be towards its natural extension to obtain star exponential functions for field theories \cite{DQC,DQF}, quantum gravity \cite{PQDQ,QPLQC,SPLQC} and, also, for the case of gauge theories for which star exponentials of the constraints are ubiquitous to achieve a gauge invariant representation of quantum mechanics in phase space \cite{SGA}. This last issue will be worked elsewhere.

\section*{Acknowledgments}
The authors would like to acknowledge support from SNII CONAHCYT-Mexico. JBM thanks to the Dipartimento di Fisica ``Ettore Pancini" for the kind invitation and its generous hospitality, and also thanks Cosimo Stornaiolo for stimulating discussions.

\section*{References}

\bibliographystyle{unsrt}

\end{document}